\documentclass[journal=jpccck,manuscript=article,layout=twocolumn]{achemso}

\usepackage{chemformula} 
\usepackage[T1]{fontenc} 
\usepackage{amsmath}
\usepackage{amssymb}
\usepackage{dblfloatfix}

\author{Christoph Schiel}
\affiliation{ 
Universit\"at Osnabr{\" u}ck, 
Fachbereich Physik, 
Barbarastra{\ss}e 7, 
D-49076 Osnabr\"uck, 
Germany}

\author{Maximilian Vogtland}
\affiliation{Fakult\"at f\"ur Chemie, Universit\"at Bielefeld, 
Universit\"atsstra{\ss}e 25,
D-33615 Bielefeld,
Germany}

\author{Ralf Bechstein}
\affiliation{Fakult\"at f\"ur Chemie, Universit\"at Bielefeld, 
Universit\"atsstra{\ss}e 25,
D-33615 Bielefeld,
Germany}

\author{Angelika K\"uhnle}
\email{kuehnle@uni-bielefeld.de}
\affiliation{Fakult\"at f\"ur Chemie, Universit\"at Bielefeld, 
Universit\"atsstra{\ss}e 25,
D-33615 Bielefeld,
Germany}

\author{Philipp Maass}
\email{maass@uni-osnabrueck.de}
\affiliation{ 
Universit\"at Osnabr{\" u}ck, 
Fachbereich Physik, 
Barbarastra{\ss}e 7, 
D-49076 Osnabr\"uck, 
Germany}

\title{Molecular self-assembly:\\ Quantifying the balance between
  intermolecular attraction and repulsion from distance and length
  distributions}

\SectionNumbersOn 
\begin{document}
\begin{abstract}
Molecular self-assembly on surfaces constitutes a powerful method for
creating tailor-made surface structures with dedicated
functionalities. Varying the intermolecular interactions allows for
tuning the resulting molecular structures in a rational fashion. So
far, however, the discussion of the involved intermolecular
interactions is often limited to attractive forces only. In real
systems, the intermolecular interaction can be composed of both,
attractive and repulsive forces. Adjusting the balance between these
interactions provides a promising strategy for extending the
structural variety in molecular self-assembly on surfaces. This
strategy, however, relies on a method to quantify the involved
interactions.\\ Here, we investigate a molecular model system of
3-hydroxybenzoic acid molecules on calcite (10.4) in ultrahigh
vacuum. This system offers both anisotropic short-range attraction and
long-range repulsion between the molecules, resulting in the
self-assembly of molecular stripes. We analyze the stripe-to-stripe
distance distribution and the stripe length distribution and compare
these distributions with analytical expressions from an anisotropic
Ising model with additional repulsive interaction. We show that this
approach allows to extract quantitative information about the strength
of the attractive and repulsive interactions. \\ Our work demonstrates
how the detailed analysis of the self-assembled structures can be used
to obtain quantitative insight into the molecule-molecule
interactions.
\end{abstract}

\newpage
\section{Introduction}
\label{intro}
Molecular self-assembly has attracted great attention due to the
impressive structural and functional variability that can be achieved
with this versatile bottom-up method for supramolecular material
synthesis \cite{RN2582}. A clever design of the molecular building
blocks allows controlling the resulting structures and tailoring them
to the specific needs of a given application \cite{RN6598}. The
interaction of the molecules with the surface provides an additional
way to tune the molecular structure formation
\cite{RN5010,RN15272,RN15275,RN13705}.

In the last decades, the subtle balance between intermolecular and
molecule-surface interactions has been explored to arrive at an
impressive multitude of various structures, ranging from perfectly
ordered two-dimensional films \cite{RN15345} over uni-directional rows
\cite{RN4367,RN11293} to porous networks,\cite{RN15268,RN6383} and
complex guest-host architectures
\cite{RN4230,RN4848,RN15269,RN15270,RN15271}. The vast majority of
these studies has focused on attractive molecule-molecule interactions
such as hydrogen bonds, van-der-Waals forces, $\pi-\pi$ interactions
or electrostatics \cite{RN6634}. In contrast, repulsive
molecule-molecule interactions have only rarely been studied for
steering the structure formation
\cite{RN15266,RN11293,RN13151,RN15273,RN15274,RN14315,RN15104,RN15276}. In
the latter examples, the electrostatic repulsion between permanent as
well as adsorption-induced electrical dipoles has been discussed as a
promising way to enhance the structural complexity in molecular
self-assembly on surfaces. Intermolecular repulsion gives rise to the
formation of homogeneously dispersed individual molecules
\cite{RN15266,RN13151}, extended rows with well-defined row-to-row
distances \cite{RN11293,RN14315} as well as islands \cite{RN15276} and
clusters \cite{RN15254} with well-defined sizes.

So far, however, the interplay between attractive and repulsive
interactions on the molecular structure formation has barely been
explored as a powerful strategy to control both the shape and the size
of self-assembled molecular structures on surfaces \cite{RN15276}. For
systematic exploring the balance between molecular attraction and
repulsion in molecular self-assembly, it is essential to
quantification the involved interactions.

Here, we present a molecular model system of adsorbed 3-hydroxybenzoic
acid molecules on a calcite (10.4) surface that provides both,
anisotropic attraction and repulsion. For this system, the molecular
self-assembly has been shown to be governed by the balance between
short-range intermolecular attraction and long-range intermolecular
repulsion \cite{RN14315,RN15104}. This balance results in the
formation of molecular stripes with a coverage-dependent
stripe-to-stripe distance distribution \cite{RN14315}.

In order to determine the strength of the involved attractive and
repulsive interactions, we consider an anisotropic Ising model with
additional long-range dipole-dipole interaction. This model is
generally applicable to stripe formation induced by intermolecular
interaction. Based on a mean-field treatment we derive analytical
expressions for stripe-to-stripe distance and stripe length
distributions.

The theory is compared with experimental data obtained by atomic force
microscopy images. An analysis of these images yields
coverage-dependent stripe-to-stripe distance distributions as well as
stripe length distributions. By fitting the theoretical predictions to
the distance and length distributions we extract the strength of the
attractive and repulsive molecule-molecule interactions. Our work
constitutes an example of how the mesoscopic structural information
can be used for gaining quantitative molecular-level insights into the
driving forces at play.

\begin{figure*}[b!]
\centering
\includegraphics[width=504pt]{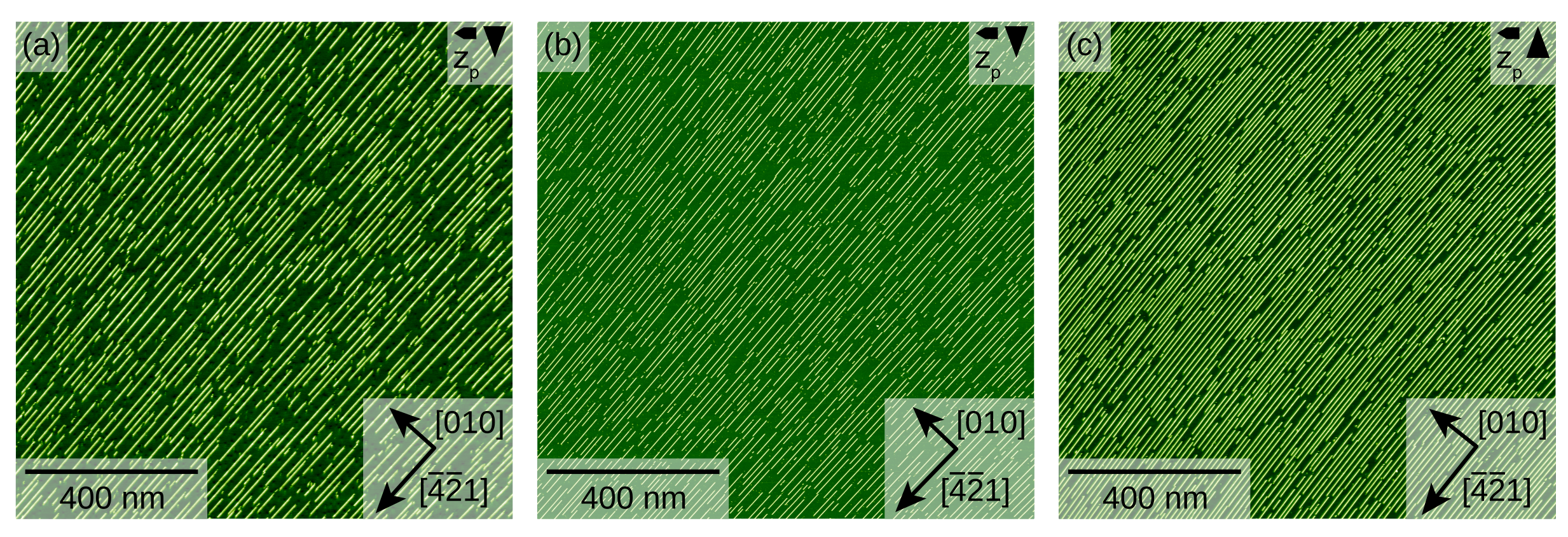}
\caption{Representative atomic force microscopy (AFM) topography
  (z$_p$) images of 3-hydroxybenzoic acid (3-HBA) on calcite (10.4)
  from the three measured series I, II, and III at temperature 290\,K
  and coverages (a) $\theta_{\rm \scriptscriptstyle I} =0.08$\,ML, (b)
  $\theta_{\rm \scriptscriptstyle II} =0.11$\,ML, and (c) $\theta_{\rm
    \scriptscriptstyle III} =0.16$\,ML. All images are cutouts with a
  size of 1150\,$\times$\,1150\,nm$^{2}$ and a resolution of
  3446\,$\times$\,3446\,Px.  The fast (small arrow) and slow (large
  arrow) scan directions are given in the upper right corner.  The
  surface directions are indicated by the arrows in the lower right
  corner.}
\label{fig:AFM_images}
\end{figure*}

\section{Methods}
\label{methods}
All dynamic atomic force microscopy (AFM) images shown in this work were acquired
with a variable-temperature atomic force microscope (VT-AFM XA from
ScientaOmicron, Germany) operating under ultrahigh vacuum conditions
($p < 10^{-11}$ mbar). We used silicon cantilevers purchased from
NanoWorld (Neuch\^atel, Switzerland) with an eigenfrequency of around
300 kHz (type PPP-NCH).  To remove contaminations and a possible oxide
layer, the cantilevers were sputtered with Ar$^+$ at 2\,keV for
10\,minutes prior to use. \\ The calcite crystals (Korth Kristalle
GmbH, Germany) were prepared \textit{ex situ} by mild ultrasonication
in acetone and isopropanol for 15\,min each. Inside the chamber, the
crystals were degassed at about 580\,K for 2\,h. After this degassing
step, the crystals were cleaved and annealed at about 540\,K for
1\,h. The quality of the crystal surface was then checked by
collecting an image of typically 100\,nm$^2$ size.  \\ The
3-hydroxybenzoic acid (3-HBA) molecules (99\,\% purity) were purchased
from Sigma-Aldrich and used after degassing for 10\,min at a
temperature higher than 320\,K. A home-built Knudsen cell with a glass
crucible was used for sublimation.  For the crucible used here, a
temperature of 309\,K resulted in a flux of approximately 0.01
monolayers per minute (ML/min). During sublimation, the partial
pressure in the chamber was in the range of $1 \times 10^{-12}$ mbar
for 3-HBA ($m/z = 137$ u/e) as measured with a mass spectrometer from
MKS (e-Vision 2).  For molecule deposition, the calcite sample was
cooled to a temperature below 220\,K.

The AFM measurements were performed at a sample temperature
of 290\,K.\footnote{In the AFM, the temperature is read out at the
  sample stage using a Pt100 sensor 3\,cm apart from the
  sample. According to the manufacturer, the temperature difference
  between the sample and the sample readout position in the AFM is
  smaller than 10\,K.} This temperature is chosen such that the
dynamics are fast enough to ensure thermodynamic equilibrium but slow
enough to minimize effects on the statistical analysis. The images
were acquired with a pixel resolution of
4000\,$\times$\,4000\,Px and a speed of 0.32\,ms/Px,
resulting in a measurement time of roughly 3\,h/image.  The image size
was 1500\,$\times$\,1500\,nm$^{2}$, yielding a
resolution of 0.375\,$\times$\,0.375\,nm$^{2}$/Px. 

We present measurement series for three different coverages, with multiple
images measured at the same location. The number of
images per coverage in each series differs since we had to sort out some of
the images due to experimental difficulties.  The remaining 14 images
resulted in total amounts of 1758254
stripe-to-stripe distances $d$ and 17015 stripe lengths $l$.

To obtain the stripe-to-stripe distance and the length distributions
from the AFM images we proceeded as follows.
After a plane subtraction and line-by-line
correction,\cite{RN12729} the images were calibrated and corrected for
linear drift \cite{Rahe.2010}.  Each image was segmented using a
trainable machine learning tool \cite{ArgandaCarreras.2017}.
Afterwards neighboring pixels were connected and the connected
structures fitted with a rectangle \cite{Legland.2016,
  Schindelin.2012}.  All relevant data of the fit rectangles (centroid
position, length $l$ and orientation) were collected and reconstructed
as line segments for further analysis using the package SpatStat
within the software R \cite{RN15282, Baddeley.2016}.  We sort out
stripes shorter than 5\,nm since these are difficult to distinguish
from wrongly fitted structures.  For simplicity, we do not exclude
stripes limited by image edges.  We define the stripe-to-stripe
distance as the distance between each 3-HBA dimer and its
next-neighbor in [010] direction.  Thus, we get one distance per
molecular dimer but only one length per stripe, which implies that the
number of measured stripe distances is much larger than the number of
stripe lengths.

\section{Experimental Results}
\label{sec:expresults}
When depositing 3-HBA molecules onto the (10.4) surface of calcite
kept in ultrahigh vacuum, the molecules self-assemble into double-rows
as has been reported previously \cite{RN14315}.  The molecular
double-rows can be identified in AFM images as stripes oriented along
the $[\overline{42}1]$ direction of the calcite crystal, see
Figure~\ref{fig:AFM_images}.  Two molecules, one out of each row,
form the stripe basis with a periodicity of 0.8\,nm \cite{RN14315}.
We call this basis a 3-HBA dimer.  Each image in
Figure~\ref{fig:AFM_images} is a representative example from one of three series I-III 
of measurements at a given coverage, where
 $\theta_{\rm \scriptscriptstyle I}$ = 0.08\,ML
[Figure~\ref{fig:AFM_images}(a)], $\theta_{\rm \scriptscriptstyle II}$
= 0.11\,ML [Figure~\ref{fig:AFM_images}(b)] and $\theta_{\rm
  \scriptscriptstyle III}$ = 0.16\,ML
[Figure~\ref{fig:AFM_images}(c)].

\begin{figure}[h!]
\centering
\includegraphics[width=0.8\columnwidth]{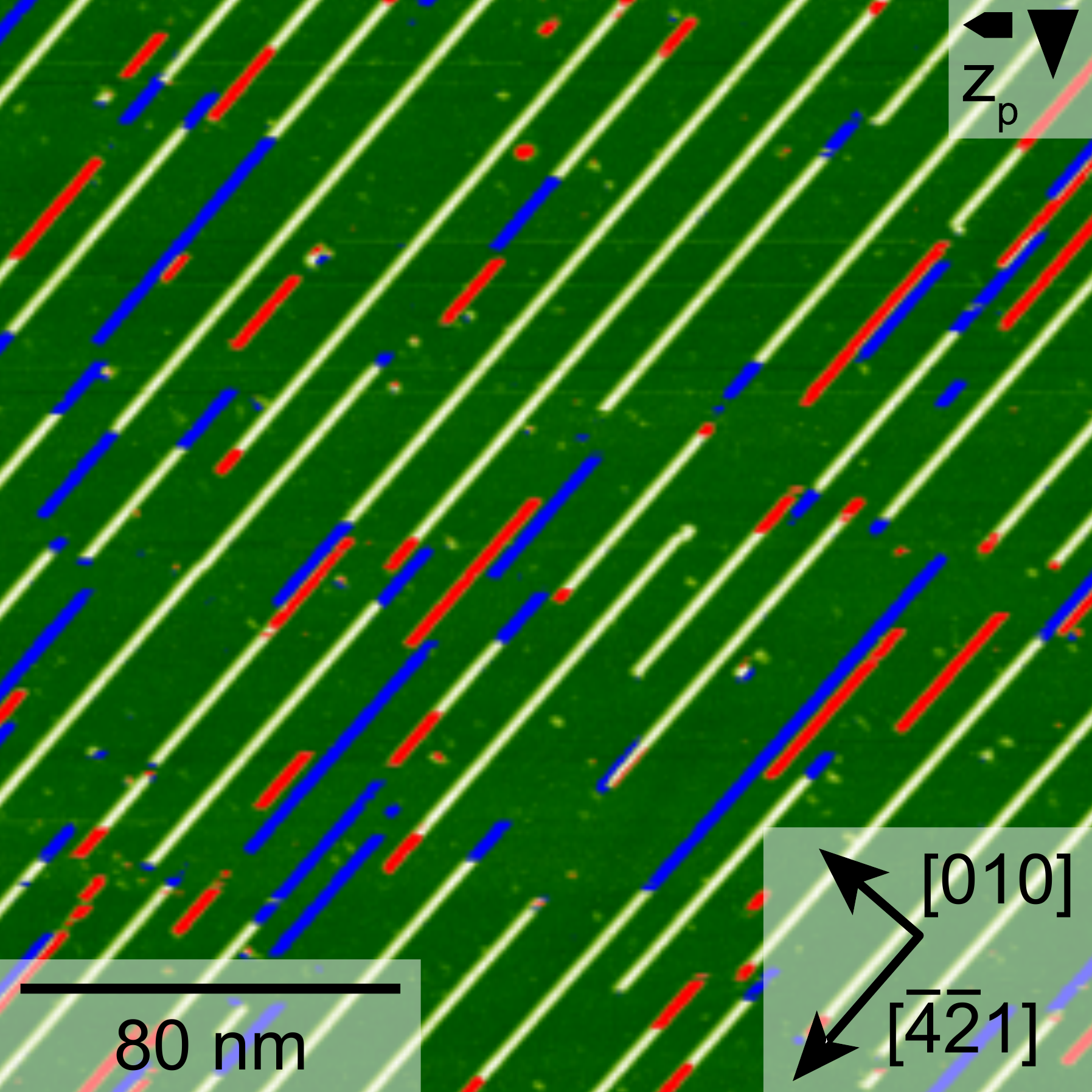}
\caption{\label{fig:diff_image} Comparison of image shown in
  Figure~\ref{fig:AFM_images}\,(b) and an image taken six hours
  before, demonstrating the redistribution of molecules. Areas where
  molecules disappear (appear) are marked in blue (red).  }
\end{figure}

In Figure~\ref{fig:diff_image}, the difference between the image shown
in Figure~\ref{fig:AFM_images}\,(b) and an image
taken six hours before is shown. The areas marked in blue (red) are regions of
disappearing (appearing) molecules over time. From this difference
image, it becomes evident that the molecules are mobile at a sample
temperature of 290\,K. More specifically, a total of $\approx$ 30\% of
the molecules, including entire stripes, change position within the
measurement time of six hours, while about 70\% of the
structures do not change.  We can thus
expect that the statistics of a single image is not strongly affected by the
long measurement time of roughly 3\,h for a single image.

\begin{figure*}[t!]
\centering
\includegraphics[width=160mm]{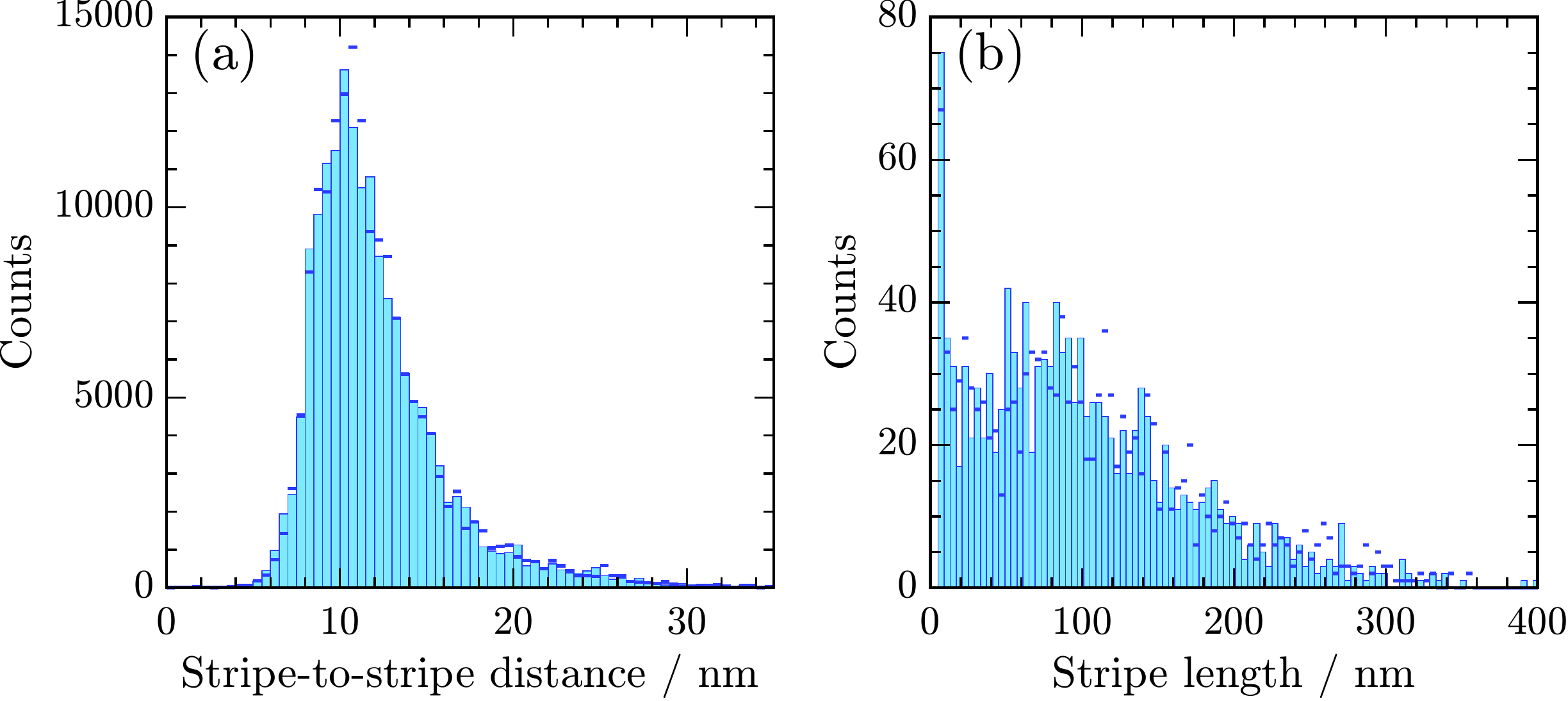}
\caption{\label{fig:diff_hists} Comparison of histograms obtained from
  the first and last image of the series III. The images have a time
  separation of 18 hours. In (a) the counts of stripe-to-stripe
  distances in bins of size 0.5\,nm are shown, and in (b) the counts
  of stripe lengths in bins of size 4\,nm. The bins give the
  histograms obtained from the first image, and the horizontal bars
  marked in blue indicate the corresponding counts from the last
  image.  }
\end{figure*}

\begin{figure*}[b!]
\centering
\includegraphics[width=160mm]{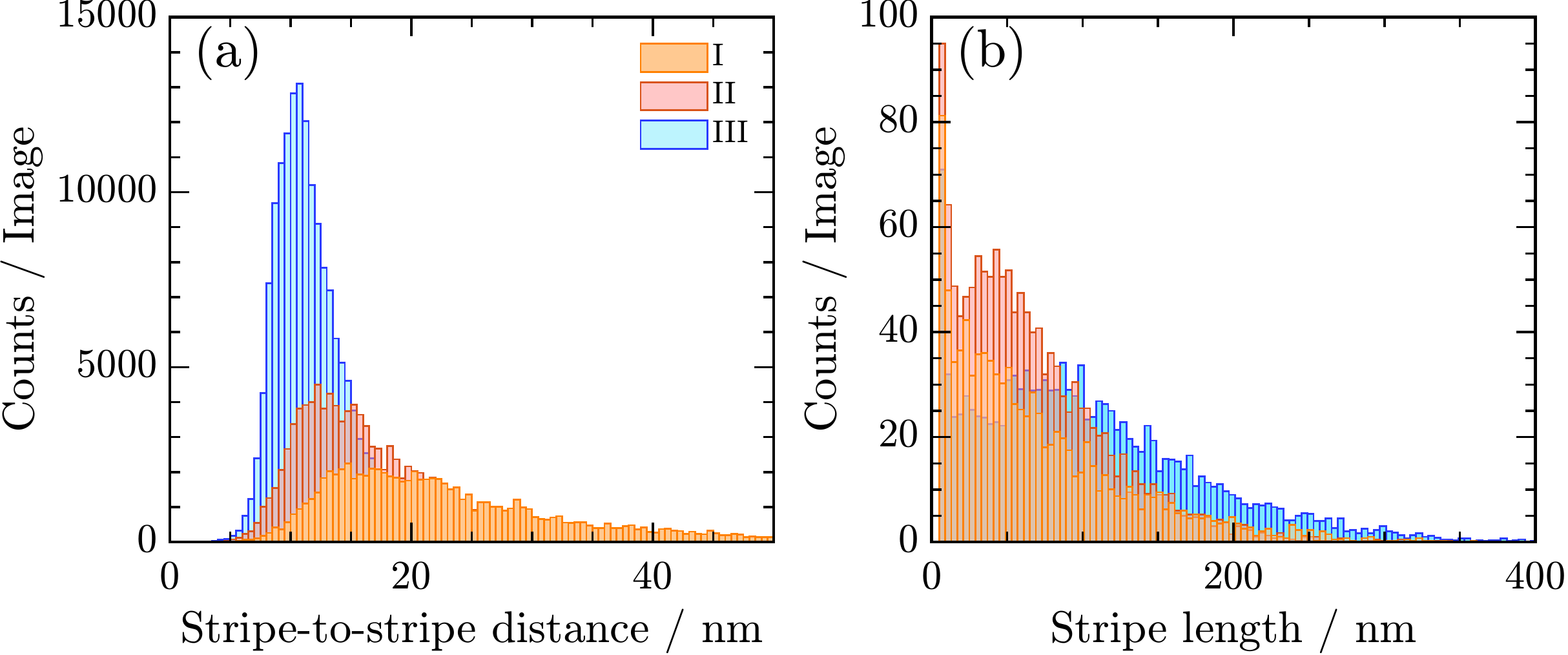}
\caption{\label{fig02} Histograms of (a) stripe-to-stripe distances
  and (b) stripe lengths obtained from all images in the series I
  (0.08 ML, 4 images), II (0.11 ML, 4 images), and III (0.16 ML, 6
  images) [color coding according to legend in (a)]. The bin sizes are
  as in Figure~\ref{fig:diff_hists}.}
\end{figure*}

From the last and the first image of series III with a time difference
of 18 hours, we have generated the stripe-to-stripe distance distributions shown
in Figure~\ref{fig:diff_hists}(a),
using a bin size of 0.5\,nm. Comparing these two distributions reveals no significant
difference.  Both distributions exhibit a distinct maximum at a distance of 10 to 12\,nm, implying
that the stripes are not randomly placed on
the surface. A random placement would result in a geometric distribution
\cite{RN14315}.  In addition have determined the stripe length distributions
for the two images, which are shown in Figure~\ref{fig:diff_hists}\,(b) for a bin size of
4\,nm.  Again the comparison of the respective two length distributions yields
no significant difference. 

To conclude, during 18 hours of measurement time appreciable rearrangements of the molecules occur, but 
the stripe-to-stripe and the length distributions do not change. Hence the stripe patterns can be regarded to 
reflect equilibrium structures. This justifies to analyze all images of each measurement series to improve the statistics.

In Figure~\ref{fig02}(a) we show the stripe-to-stripe distance 
obtained from all images in each series (four images for
series I and II, and six images for series III). 
These distributions are coverage-dependent,\cite{RN14315} exhibiting
a decrease of mean distance ($\bar{d}_{\rm \scriptscriptstyle I}$ =
24.1\,nm, $\bar{d}_{\rm \scriptscriptstyle II}$ = 18.2\,nm and
$\bar{d}_{\rm \scriptscriptstyle III}$ = 12.2\,nm), standard deviation
($\sigma_{\rm \scriptscriptstyle I}$ = 12.2\,nm, $\sigma_{\rm
  \scriptscriptstyle II}$ = 8.5\,nm and $\sigma_{\rm
  \scriptscriptstyle III}$ = 4.0\,nm) and position of the maximum with
increasing coverage.

The corresponding length distributions are shown in Figure~\ref{fig02}(b). As
explained above, the number of counts in each bin is much less than
for the stripe distances. Overall, the length distributions decrease
monotonically for large $l$. For the two higher coverages (series II and III), local
maxima in the range $l\,{\approx}\,50{-}100$\,nm 
appear. A corresponding maximum, however, is not clearly detectable at
the lowest coverage (series I).

The total numbers of stripes are $N_{\rm \scriptscriptstyle I}$ =
959\,stripes/image, $N_{\rm \scriptscriptstyle II}$ =
1370\,stripes/image, $N_{\rm \scriptscriptstyle III}$ =
1284\,stripes/image in series I-III.
We determined mean lengths $\bar{l}$ ($\bar{l}_{\rm \scriptscriptstyle
  I}$ = 70.3\,nm, $\bar{l}_{\rm \scriptscriptstyle II}$ = 65.8\,nm,
$\bar{l}_{\rm \scriptscriptstyle III}$ = 104.4\,nm) and respective standard
deviations $\sigma_{l}$ ($\sigma_{l,\rm \scriptscriptstyle I}$ =
59.6\,nm, $\sigma_{l,\rm \scriptscriptstyle II}$ = 49.0\,nm,
$\sigma_{l,\rm \scriptscriptstyle III}$ = 92.2\,nm) for the three
series.  While we can see and expect a global trend of increasing mean
length and standard deviation with increasing coverage, both values
are smaller for series II compared to series I.

\section{\label{sec:model} Theoretical modeling}
For the equilibrated system of 3-HBA molecules on calcite there it has
been proposed that repulsive interactions are caused by a charge
transfer between surface and molecules, leading to dipolar moments
perpendicular to the surface \cite{RN14315}. As the stripe formation
is formed by dimers, it is convenient to consider these as molecular
units occupying lattice sites. We refer to them as ``particles''. The
lattice sites correspond to the anchoring positions on the calcite
surface.

The analysis of AFM images shows that the stripes have a width of 2~nm
and a periodicity of 0.8~nm \cite{RN14315}. This can be represented by
a rectangular lattice with spacings $a_{\parallel} = 0.8 a_0$ in
stripe direction and $a_{\perp} = 2 a_0$ perpendicular to it, where
$a_0=1\,\textrm{nm}$ sets our length unit.

The interplay between attractive and dipolar interactions is described
by the lattice gas Hamiltonian
\begin{equation} \label{eq:latticegashamilton}
H = -\frac{J}{2}\sum_{i \,{\rm NN}\,j }n_in_j + \frac{\Gamma}{2}\sum_{k,l}\frac{n_k n_l}{r_{kl}^3} \,.
\end{equation}
Here $n_i$ are occupation numbers, \textit{i.e.}, $n_i = 1$ if the
site $i$ is occupied by a particle and zero otherwise. The sum over
$i$ and $j$ is restricted to nearest-neighbor (NN) sites in stripe
direction corresponding to an anisotropic Ising model, and $r_{kl}$ is
the (dimensionless) distance between sites $k$ and $l$. The
interaction parameter $J>0$ quantifies the strength of the attractive
nearest-neighbor interaction. The strength of the repulsive dipole
interaction is given by
\begin{equation}
\Gamma=\frac{p^2}{4\pi\epsilon_0a_0^3}\,,
\end{equation}
where $p$ is the dipole moment of one dimer and $\epsilon_0$ is the dielectric permeability of the vacuum. 

In the following two subsections, we discuss analytical approaches to
get insight into equilibrated stripe patterns for $\Gamma = 0$, and
for $\Gamma>0$ based on approximate one-dimensional treatments. This
allows one to determine the interaction parameters $J$ and $\Gamma$ by
fitting analytical expressions to match experimentally observed stripe
distance and length distributions. For convenient notation in the
following theoretical treatment, the stripe length $l$ is given in
units of $a_\parallel$ and the stripe distance $d$ in units of
$a_\perp$.

\subsection{Stripe formation for $\Gamma=0$}
\label{subsec:g0}
In the absence of dipolar interactions, the stripe positions in
perpendicular direction are uncorrelated. As a consequence, the stripe
distance distribution $\Phi_0(d)$ is geometric, ${\Phi_0(d) = \theta
  (1-\theta)^{d-1}}$.

For deriving the stripe length distribution, we can focus on a
one-dimensional row of stripes.  A stripe of length $l$ corresponds to
an occupation number sequence $01\ldots10$, \textit{i.e.}, a
configuration of two zeros separated by $l$ ones.  We denote the
probability of such sequence by $q_l$. Knowing $q_l$, the stripe
length distribution is $\psi(l)=q_l/\sum_{l=1}^\infty q_l$.

To determine $q_l$, we introduce the conditional probabilities
$w(n_{i+1}|n_i,n_{i-1},\ldots, n_1)$ of finding occupation number
$n_{i+1}$ if the occupation numbers $n_{i},n_{i-1}\dots n_{1}$ are
given. In the grand-canonical ensemble these satisfy the Markov
property $w(n_{i+1}|n_i,n_{i-1},\ldots, n_1) = w(n_{i+1}|n_i)$
\cite{Buschle/etal:2000}. Accordingly, $q_l=(1-\theta)
w(1|0)w(1|1)^{l-1}w(0|1)$, where the factor $(1-\theta)$ accounts for
the first zero in the sequence, and the product of $w(.|.)$ is the
Markov chain corresponding to the occupation numbers in the sequence.
The conditional probability $w(1|1)$ is given by $w(1|1) =
\chi_{2}(1,1)/\chi_{1}(1)$ with $\chi_{1}(1) = \theta$, and the joint
probability $\chi_{2}(1,1)$ is equal to the equilibrium
nearest-neighbor correlator \cite{Dierl/etal:2013}
\begin{align}  
C(J)&= \langle n_i n_{i+1} \rangle_{\textrm{eq}} \nonumber \\
&= \theta +\frac{1
- \sqrt{1+4\theta(1-\theta)(e^J - 1)}}{2(e^J-1)} \,.
\label{eq:correlator}
\end{align}
Hence, the $l$ dependence of $q_l$ is $\propto (C/\theta)^l$, and for the length distribution we obtain
\begin{equation} 
\Psi_0(l) = \frac{\theta-C(J)}{C(J)} \left(\frac{C(J)}{\theta}\right)^{l} \,,
\label{eq:psil}
\end{equation}
in agreement with results earlier reported in
Ref.~\cite{Yilmaz/Zimmermann:2005}{.} For $J\to0$, $C(0)=\theta^2$ and
we obtain the geometric distribution
$\Psi_0(l)=(1-\theta)\theta^{l-1}$.

\subsection{Stripe formation for $\Gamma>0$}
\label{subsec:gl0}

The dipolar interaction for $\Gamma>0$ leads to repulsion between
pairs of particles belonging to the same stripe as well as to
different stripes. It tends to shorten the stripes and to increase the
stripe distances. Compared to the case $\Gamma=0$, the stripe distance
distribution is more strongly affected than the length distribution,
because the latter is largely determined by the attractive
nearest-neighbor interaction $J$ (if $\Gamma < J$~).

In fact, one can expect that the length distribution for large $l$ is
still geometric as in Eq.~\eqref{eq:psil} for $\Gamma=0$.  This is
because for each $\Gamma>0$ there is a characteristic length scale of
induced correlations by the dipolar interaction. Considering long
stripes to be composed of particle blocks with this length scale, the
reasoning in the previous subsection leading to Eq.~\eqref{eq:psil} is
applicable with a renormalized $C = C_{\rm eff}(J,\Gamma)$ in
Eq.~\eqref{eq:correlator}. Hence, the length distribution in the
presence of dipolar interactions is expected to decay exponentially
for large $l$ and to show deviations from the geometric shape at small
stripe lengths.

Exact analytical solutions for the distance and length distributions
are not available in the presence of competing attractive
nearest-neighbor and dipolar interactions. We therefore rely on
approximate treatments here.

As for the stripe lengths, it is instructive to first analyze whether
a single isolated stripe can have an energetic minimum at a finite
length.  When increasing the length of this single stripe from $l$ to
$l+1$, the energy changes by
\begin{equation} \label{eq:singlestripe}
	\Delta H(l) = -J + \Gamma \sum_{k=1}^{l}\frac{1}{k^3} \,.
\end{equation} 
For a minimum to occur, $\Delta H(l)$ must be negative for $l=1$ and
positive for $l\rightarrow\infty$. This implies
$1<J/\Gamma<\zeta(3)\cong1.202$, where $\zeta(3)$ is the Riemann zeta
function (Ap\'ery's constant). Accordingly, a finite single stripe can
form only in a narrow regime of the interaction parameters $J$ and
$\Gamma$. However, in a system of many interacting stripes at a given
coverage, the stripes can mutually stabilize each other at finite
lengths for a wide range of $J$ and $\Gamma$.

Due to the fast convergence of the sum in Eq.~\eqref{eq:singlestripe},
the energy change $\Delta H(l)$ for attaching one further particle to
a stripe becomes essentially constant for $l \gtrsim 10$. We thus can
expect Eq.~\eqref{eq:psil} to hold for large $l$ with $C_{\rm
  eff}(J,\Gamma)=C(J_{\rm eff})$, where
\begin{equation}
J_{\rm eff} = J - \zeta(3) \Gamma \,.
\label{eq:jeff}
\end{equation}
The corresponding approximate stripe length distribution is referred to as $\tilde{\Psi}(l)$. 

We expect this distribution to have the same asymptotic behavior as the true length distribution $\Psi(l)$, \textit{i.e.}
\begin{equation} \label{eq:tildepsi}
\Psi(l) \sim \tilde{\Psi}(l) \sim \bigg(\frac{C(J_{\rm eff})}{\theta}\bigg)^l \,, \quad l \to \infty \,.
\end{equation}
Deviations from $\tilde{\Psi}(l)$ are expected to be significant for
small $l$. If the effective nearest-neighbor interaction $J_{\rm eff}$
is attractive, {\it i.e.}, $J>\zeta(3)\Gamma$, the energy change
$\Delta H(l)$ in Eq.~\eqref{eq:singlestripe} is negative, implying
that single particles or small stripes are energetically unfavorable
compared to longer stripes. Accordingly, we expect $\Psi(l)$ to be
smaller than $\tilde{\Psi}(l)$ for small $l$.

As for the stripe distance distribution $\Phi(d)$, we can assume that
it is governed by the dipolar interaction between neighboring stripes
in perpendicular direction.  Applying a mean-field approach similar to
that introduced in Ref.~\cite{RN14315}, we divide the two-dimensional
stripe pattern into mutually independent one-dimensional parallel
bands in perpendicular direction. The bands are considered to have the
same width $\bar{l}$, where $\bar{l}$ is the mean stripe lengths.

For each stripe appearing in a band, we consider it to span the whole
band, \textit{i.e.}\ to have length $l=\bar{l}$. In one band, the
interaction $U(d)$ between two stripes at distance $d$ with dipole
density $p/a_{\parallel}$ is (integrating along both stripes with
parametrization $s_1$ and $s_2$)
\begin{align} \label{eq:ud}
U(d) &= \frac{p^2}{4\pi\epsilon_0a_\parallel^2}
\int\limits_{-\bar{l}/2}^{\bar{l}/2}\hspace{-0.3em}ds_1 \hspace{-0.3em}
\int\limits_{-\bar{l}/2}^{\bar{l}/2} \hspace{-0.3em}ds_2\, \frac{1}{|\vec{x}_1(s_1) - \vec{x}_2(s_2)|^3} \nonumber \\ 
&= \frac{p^2}{2\pi\epsilon_0da_\parallel^2}\Biggl[\Bigl(1+\frac{\bar{l}^{\,2}}{d^2}\Bigr)^{1/2}-1\Biggr] \,.
\end{align}
Hence, we have mapped each band onto a one-dimensional lattice
occupied by particles with interaction $U(d)$ between neighboring
stripes.

The mean occupation of lattice sites is fixed by the coverage
$\theta$. In the presence of the purely repulsive $U(d)$, it can be
viewed as resulting from a confinement pressure $f$ which hinders the
particles to become infinitely separated and to give rise to a mean
distance $\bar d$.  Our approximation $\tilde{\Phi}(d)$ of the stripe
distance distribution thus is given by
\begin{subequations}
\label{eq:tildephi}
\begin{equation} 
\label{eq:tildephi}
\tilde{\Phi}(d)=\frac{1}{Z}\text{exp}(-\beta[fd+U(d)])\,,
\end{equation}
where $Z= \sum_{d=1}^{\infty}\text{exp}(-\beta[fd+U(d)])$ and $f$ is
fixed by the condition
\begin{equation} \label{eq:tildephi_c}
\bar{d} = \sum_{d=1}^{\infty}d\tilde{\Phi}(d)\,.
\end{equation}
\end{subequations}

\begin{figure}[tp]
\includegraphics[width=\columnwidth]{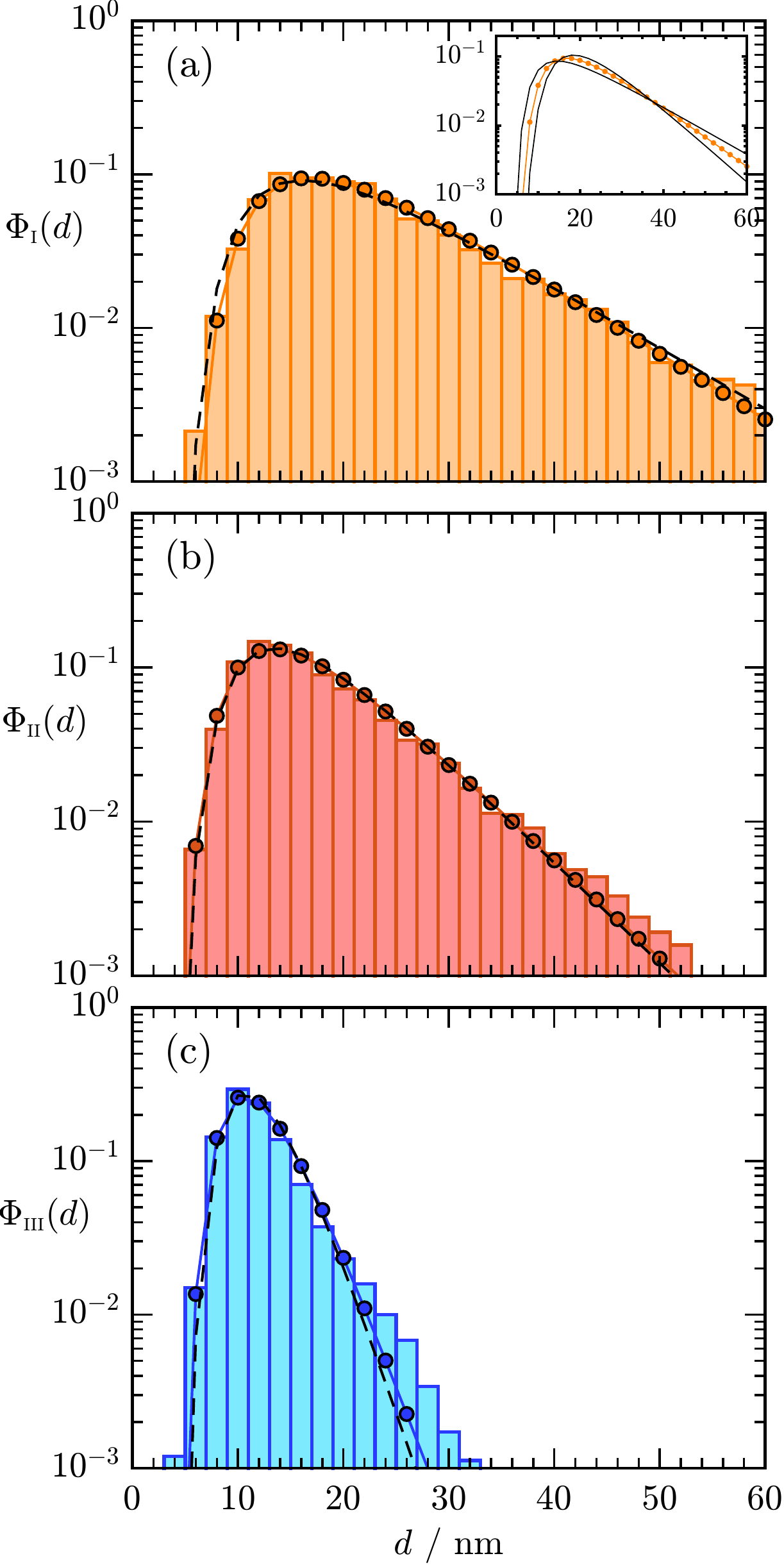}
\caption{Histograms of the measured distance distributions $\Phi(d)$
  for the three different coverages (a) $\theta_{\rm
    \scriptscriptstyle I} = 0.08 \,\textrm{ML}$, (b) $\theta_{\rm
    \scriptscriptstyle II} = 0.11\, \textrm{ML}$, and (c) $\theta_{\rm
    \scriptscriptstyle III}=0.16\, \textrm{ML}$ in comparison with the
  fitted theoretical distributions $\tilde{\Phi}(d)$ (circles,
  connected by solid lines). Dashed black lines correspond to
  $\tilde{\Phi}(d)$ with the mean dipole moment of $\bar{p} =
  6.3~\textrm{D}$. The inset in (a) shows the fitted $\tilde{\Phi}(d)$
  for $\theta_{\rm \scriptscriptstyle I}$ (circles, connected by
  orange line) compared to $\tilde{\Phi}(d)$ for $p_> =
  8.3~\textrm{D}$ and $p_<=4.3~\textrm{D}$ (black lines).}
\label{fig:DistanceFit}
\end{figure}

\section{Application to experiments}
The parameters $J$ and $\Gamma$ are estimated by fitting
$\tilde{\Phi}(d)$ from Eq.~\eqref{eq:tildephi} to the distribution
$\Phi(d)$, and by fitting Eq.~\eqref{eq:tildepsi} to the tail of
$\Psi(l)$, where $\Phi(d)$ and $\Psi(l)$ are the distributions
obtained in the experiments.

We first determine $\Gamma$, and hence $p = \sqrt{\Gamma 4\pi\epsilon_0a_0^3}$, by fitting  $\tilde{\Phi}(d)$ to $\Phi(d)$ with the experimental $\bar{l}$ in Eq.~\eqref{eq:ud}. We then extract $J_{\rm eff}$ by fitting the tail of $\Psi(l)$ which yields $J$ via Eq.~\eqref{eq:jeff}.

\begin{table}[t]
\renewcommand{\arraystretch}{1.3}
\begin{tabular}{|c|c|c|c|c|}
\hline 
Series & $\theta$ / ML & $p$ / D & $\Gamma$ / meV & $J$ /eV \\
\hline
I& 0.08 & 7.0 & 31 & 0.32  \\
\hline
II& 0.11 & 6.1 & 24 & 0.28 \\
\hline
III& 0.16 & 5.8 & 21 & 0.28  \\
\hline
\end{tabular}
\caption{Parameters obtained from fitting the theoretical model to the experimental stripe distance and length distributions for the three different coverages (series I-III).}
\label{tab:parameters}
\end{table}

Figure~\ref{fig:DistanceFit} shows fits of $\tilde{\Phi}(d)$ (circles,
connected by solid lines) to $\Phi(d)$ (histogram) for each series,
using the method of least square.  The optimal values of $\Gamma$ (and
corresponding $p$) for each coverage are listed in
Table~\ref{tab:parameters}. In all three cases, the predicted curves
match the experiment. The fitted dipole moment decreases from
$7.0\,\textrm{D}$ to $5.8\,\textrm{D}$ with increasing $\theta$. When
fixing the dipole moment to the mean $\bar{p}= 6.3\,\textrm{D}$ of
these values, the corresponding $\tilde{\Phi}(d)$ are also in good
agreement with the experiment, as shown by the dashed black lines in
Figure~\ref{fig:DistanceFit}.

The mean $\bar p$ differs by about 1\,D from the optimal value for the
smallest coverage. This raises the question on the sensitivity of the
fitting with respect to $p$. We thus analyze how $\tilde{\Phi}(d)$
deviates from $\Phi(d)$ for even larger differences of $p$ from its
optimal value. For values $p_> = 8.3\,\textrm{D}$ and $p_< =
4.3\,\textrm{D}$ larger and smaller by 2\,D, $\tilde{\Phi}(d)$ is
shown in the inset of Figure~\ref{fig:DistanceFit}(a). As can be seen
from this inset, deviations to $\tilde{\Phi}(d)$ for the optimal $p$
value are now clearly visible. We thus conclude that the error in our
estimate is about~$\pm 1$\,D.

Taking $\bar{p}$ as the dipole moment of the 3-HBA dimer yields a
dipole moment $p/2$ = 3.2\,D for the single molecule, in fair
agreement with our former estimate \cite{RN14315}.

\begin{figure}[tp]
\includegraphics[width=\columnwidth]{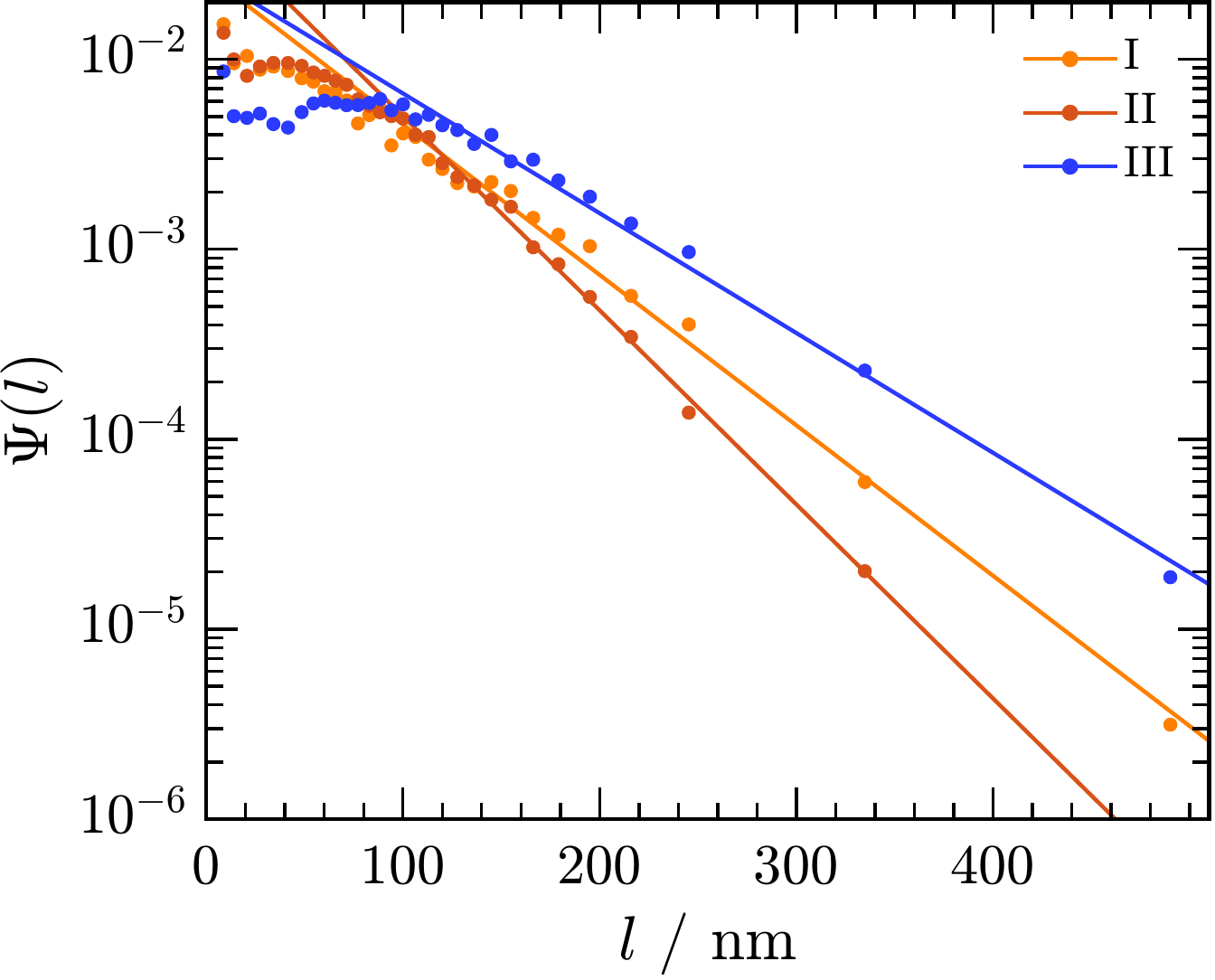}
\caption{Stripe length distributions $\Psi(l)$ for the three different
  coverages (I-III, circles) with fits to the exponential tails for
  $l>100~\textrm{nm}$ (solid lines).}
	\label{fig:lengthhists}
\end{figure}

Having determined $\Gamma$, we now analyze the stripe length
distribution to determine $J$. Figure~\ref{fig:lengthhists} shows the
measured length distributions $\Psi(l)$ for each series (circles). The
distributions are determined by using bins of varying size with
approximately equal amount of events in each bin. As expected, all
distributions show an exponential decay for large $l$. The solid lines
are fits to these exponential decays for $l>100 \, \textrm{nm}$.
According to Eq.~\eqref{eq:tildepsi}, the decay length of $\Psi(l)\sim
e^{-l/l_0}$ is
\begin{equation} \label{eq:l0}
l_0 = \frac{a_\parallel}{\ln (C(J_{\rm eff})/\theta)}\,.
\end{equation}
The characteristic decay length $l_0$ for each experimental
distribution thus yields a value $J_{\rm eff}$ via Eq.~\eqref{eq:l0}
in combination with Eq.~\eqref{eq:correlator}. The interaction
parameters $J$ then follow from Eq.~\eqref{eq:jeff} and are listed in
the fifths column of Table~\ref{tab:parameters}. These values lie
around $0.29~\textrm{eV}$. Our final estimate of the analysis is $p =
6.3~\textrm{D} \pm 1\textrm{D}$ and $J = 0.29 \pm 0.04~\textrm{eV}$.

The interaction strength $J$ is in the range of hydrogen-bonds and
lower than ${\approx}\,0.7$\,eV between two molecules in a carboxylic
acid dimer \cite{Steiner.2002,Martsinovich.2010, Tzeli.2013}.

\section{Conclusions}
In summary, we have presented an approach to estimate the strengths of
short-range attractive and long-range repulsive interactions between
3-HBA molecules on a calcite surface by an analysis of
stripe-to-stripe distance distributions $\Phi(d)$ and stripe length
distributions $\Psi(l)$.

Experimental distributions were determined from an analysis of three
AFM image series with different coverages 0.08\,ML, 0.11\,ML, and
0.16\,ML at a temperature 290\,K. The measurements of theses series
spanned time intervals of up to 18 hours. A comparison between
distributions of individual images in the same series strongly
suggests that the stripe patterns are in thermodynamical equilibrium.

The attractive interaction responsible for the stripe formation was
considered to be an effective one with strength $J$ between
neighboring 3-HBA dimers, without resorting to details of the
molecular structure. The long-range repulsive interaction is modeled
as dipole-dipole interaction of characteristic strength $\Gamma$ as
previously proposed in Ref.~\cite{RN14315}{.} It is believed to be
caused by a charge transfer between the surface and 3-HBA molecules.
As these molecules have specific anchoring sites on the calcite
surface, the system could be described by a lattice gas corresponding
to an anisotropic Ising model with additional dipolar interaction.

Based on this model, we developed mean-field approaches to derive
approximate expressions for the stripe distance and length
distributions with $J$ and $\Gamma$ as parameters. Fitting these
parameters to the experimental distributions we obtained the estimates
$J = 0.29 \pm 0.04~\textrm{eV}$ and $p = 6.3~\textrm{D} \pm
1\textrm{D}$ for the dipole moment $p \propto \sqrt{\Gamma}$ of a
3-HBA dimer.

The modeling approach presented here is applicable also to other
molecular systems self-assembling into stripe patterns, if the stripe
formation is dominated by short-range attractive molecule-molecule
interactions. In general, one can expect additional long-range
electrostatic interactions to be present. Their impact on the
structure formation depends on their type (e.g., dipolar, quadrupolar)
and strength, but the core of our methodology is independent of these
features.

The mean-field treatment, however, requires the formation of
structures with long stripes arranging into patterns with large
overlaps between neighboring parallel stripes. This requirement is
fulfilled only if the repulsive interaction is not too strong compared
to the attractive one, and if the coverage is not too small. The
coverage must not be too high either because otherwise the structure
will no longer be composed of individual stripes. For determining the
respective limits of our mean-field treatment, extensive simulations
of the many-body problem are needed, which is left for future
research.

As long as the aforementioned requirements are met, other types of
interactions can be accounted for by minor adjustments of the
mean-field approach.  As for the stripe distance distribution, only
the effective interaction potential $U(d)$ between stripes in
Eq.~\eqref{eq:ud} needs to be modified.  As for the stripe length
distribution, we expect a length scale to exist beyond which
correlations within a stripe can be renormalized to an effective
nearest-neighbor interaction between segments. The interplay between
attractive and repulsive interaction in $\Phi(l)$ can then be
accounted for by one effective coupling parameter analogous to $J_{\rm
  eff}$ in Eq.~\eqref{eq:jeff}.

From a general point of view, it should be scrutinized whether a
modeling with static dipole moment is appropriate. Our use of a static
dipole moment here relies on the assumption of an approximately fixed
amount of charged transferred between the surface and each
molecule. The results in Table~\ref{tab:parameters} indicate a
decreasing dipole moment with increasing coverage. This can be
interpreted by a dynamic dipole moment which becomes smaller in order
to compensate for additional repulsive interactions with further
molecules. A change of the molecule-surface interaction as a response
to a repulsive interaction has been reported earlier in
Ref.\cite{DellaPia.2014, Duhm.2013, Fraxedas.2011}.

Dynamical dipole moments can be coped with in a theoretical treatment
by introducing a molecular polarizability for the molecules. This
leads to varying dipole moments in dependence of their local
environment. How important these variations are, is presently
unknown. The uncertainties of the values in Table~\ref{tab:parameters}
and the rather narrow coverage range 0.08-0.16~ML does not allow us to
give a firm assessment on how strong effects of a dynamical dipole
moment are. Additional investigations with a wider range of coverages
are needed. Further experimental and theoretical research in this
direction offers promising perspectives to gain deeper insight into
the impact of the interplay between repulsive and attractive
interactions on molecular self-assembly.

\begin{acknowledgement}
We thank the German Research Foundation (DFG) for funding (KU
1980/10-1) "Intermolecular Repulsion in Molecular Self-Assembly on
Bulk Insulator Surfaces".
 
%

\end{acknowledgement}

%



\providecommand{\latin}[1]{#1}
\makeatletter
\providecommand{\doi}
  {\begingroup\let\do\@makeother\dospecials
  \catcode`\{=1 \catcode`\}=2 \doi@aux}
\providecommand{\doi@aux}[1]{\endgroup\texttt{#1}}
\makeatother
\providecommand*\mcitethebibliography{\thebibliography}
\csname @ifundefined\endcsname{endmcitethebibliography}
  {\let\endmcitethebibliography\endthebibliography}{}

\end{document}